\documentclass[%
reprint,
superscriptaddress,
showpacs,preprintnumbers,
amsmath,amssymb,
prb,
]{revtex4-2}
\usepackage{graphicx}
\usepackage{dcolumn}
\usepackage{xfrac}
\usepackage{bm}
\usepackage{color}
\usepackage[export]{adjustbox}
\usepackage{xr-hyper}
\usepackage{hyperref}
\usepackage{tabularx}
\usepackage{lipsum}

\usepackage{amsmath}
\usepackage{amsfonts}
\usepackage{amssymb}
\usepackage{tabularx}
\usepackage{chemformula}
\usepackage{soul,xcolor}
\setstcolor{red}

\newcommand{\TC}{$T_{\text C}$}

\newcommand{\ub}{$\mu_{\text B}$}

\newcommand{\JKK}{Jkg$^{-1}$K$^{-1}$}
\newcommand{\JK}{Jkg$^{-1}$}

\begin{document}

\title{Engineering the Magnetocaloric Effect in Nd$T_4$B}

\author{Kyle~W.~Fruhling}
\email{fruhling@bc.edu}
\affiliation{Department of Physics, Boston College, Chestnut Hill, MA 02467, USA}

\author{Enrique~O.~González~Delgado}
\affiliation{Department of Physics, Boston College, Chestnut Hill, MA 02467, USA}

\author{Siddharth~Nandanwar}
\affiliation{Department of Physics, Boston College, Chestnut Hill, MA 02467, USA}

\author{Xiaohan~Yao}
\affiliation{Department of Physics, Boston College, Chestnut Hill, MA 02467, USA}

\author{Zafer~Turgut}
\affiliation{Aerospace Systems Directorate, Air Force Research Laboratory, Wright Patterson Air Force Base, OH 45433, USA}

\author{Michael~A.~Susner}
\affiliation{Materials and Manufacturing Directorate, Air Force Research Laboratory, Wright Patterson Air Force Base, OH 45433, USA}

\author{Fazel~Tafti}
\email{fazel.tafti@bc.edu}
\affiliation{Department of Physics, Boston College, Chestnut Hill, MA 02467, USA}

\begin{abstract}

We present a comprehensive study of the magnetocaloric effect (MCE) in the \ch{Nd$T$4B} system where $T$ = Fe, Co, and Ni. 
These compounds are ferromagnetic kagome materials with tunable ordering temperatures, transition width, and magnetic moments depending on the choice of transition metal.
Thus, they are good candidates for investigating the MCE.
We characterize the MCE using standard metrics and construct ternary phase diagrams as functions of Fe, Co, and Ni concentrations. 
Using these phase diagrams, we engineer the composition \ch{NdFe_{1.15}Co_{0.46}Ni_{2.39}B} to maximize the MCE. 
Interestingly, the \ch{Nd$T$4B} system shows a notable entropy change over a wide temperature range ($\sim$10 to 650 K), and particular compositions have notable MCEs spanning hundreds of Kelvin, making this a suitable system to study for technologies used in a wide range of temperatures.
In a few cases, we observe a two-peak MCE. 
These two transitions, releasing comparable entropy, provide an interesting platform to study for applications in multi-stage cooling. 
\end{abstract}

\maketitle


\section{\label{sec:introduction}Introduction}

Refrigeration and air conditioning are responsible for 25\% of residential energy consumption in the United States~\cite{gutfleisch_magnetic_2011}. 
Vapor compression cooling technology, which was developed nearly a century ago, has several adverse environmental and practical impacts.
It uses atmospherically damaging gases, draws considerable electrical power, and has many mechanical parts that produce unwanted noise and vibrations and have the potential to break.
A potential solution to these issues is magnetic cooling based on the magnetocaloric effect (MCE), which offers an alternative technology that is more energy-efficient and free from noise, vibrations, and atmospherically damaging gases~\cite{gutfleisch_magnetic_2011, mellari_introduction_2023, zulnehan_magnetocaloric_2024, zhang_advanced_2024}.

The magnetocaloric effect is the reversible change of temperature and entropy of a magnetic material when subjected to a magnetic field, under adiabatic and isothermal conditions~\cite{romero_gomez_magnetocaloric_2013}.
In the 1930s, this technique was used to achieve sub-Kelvin cooling by performing cyclic isothermal magnetization and adiabatic demagnetization on a paramagnetic salt~\cite{giauque_attainment_1933}.
The giant magnetocaloric effect, found in the 1990s, is a separate phenomenon from the conventional MCE.  Here, in ferromagnetic (FM)  compounds such as \ch{Gd5Si2Ge2}, the change in magnetic entropy resulting from the magnetic transition governs the effect~\cite{pecharsky_giant_1997}. 
From a practical point of view, transitions (or, more specifically, a series of transitions) from around room temperature to low temperatures are desired, thus making systems with a tunable \TC\ an intense research target. 
Indeed, prototype systems evincing efficiencies of 20-30\% greater than standard vapor compression technologies have already been built~\cite{yu_review_2010}.

 Even with these successes, effective commercialization requires magnetocaloric materials to exhibit 1) tunable transition temperatures that span the desired cooling range with 2) concomitant large entropy changes associated with these transitions and 3) mechanical stability due to the large degree of temperature and phase cycling these materials will undergo during their lifetimes. 
 In order to tune $T_C$ to temperatures of interest, or to increase the change in entropy to the point that magnetocaloric technologies are economically viable, new material families can be investigated or previously studied families can be tuned with chemical or mechanical means~\cite{czernuszewicz_discovery_2025,ghorai_giant_2024, qiao_giant_2025,zheng_giant_2023, abramchuk_tuning_2019,fruhling_characterization_2024, caron_pressure-tuned_2011,pan_enhancement_2025,pecharsky_tunable_1997,yang_giant_2023, yang_tunable_2020,bykov_magnetocaloric_2021,zhang_tunable_2014}. 

In this work, we investigate the MCE of the highly tunable \ch{Nd$T$4B} system where $T$ = Fe, Co, and Ni. 
These materials were chosen because they exhibit FM transitions, have large magnetic moments, and are tunable based on the choice of transition metals ($T$).
The parent compound \ch{NdNi4B} has a Curie temperature, $T_\text{C}=12$~K and saturation moment $M_S=2.1$~\ub\ per formula unit.
When Ni is completely replaced with Co, the Curie temperature and saturation moment jump to 468~K and 5.8~\ub.
With the replacement of one Co atom with Fe in \ch{NdFeCo3B}, there is a further significant jump to $T_\text{C}=688$~K and $M_S=6.8$~\ub~\cite{burzo_rare_2023}.
Due to the large tunability of the moment size and Curie temperature, these materials seemed ideal for engineering the magnitude and temperature range of the MCE. 
We chose to focus on Nd-based compositions as they exhibit globally higher magnetic moments than compounds with the same formula but differing rare earth elements~\cite{burzo_rare_2023}. 
Additionally, boride materials are generally considered good candidates for industrial applications due to their chemical and structural stability~\cite{markovskii_chemical_1962, pangilinan_hardening_2022}. 

\begin{figure}
\centering
  \includegraphics[width=0.5\textwidth]{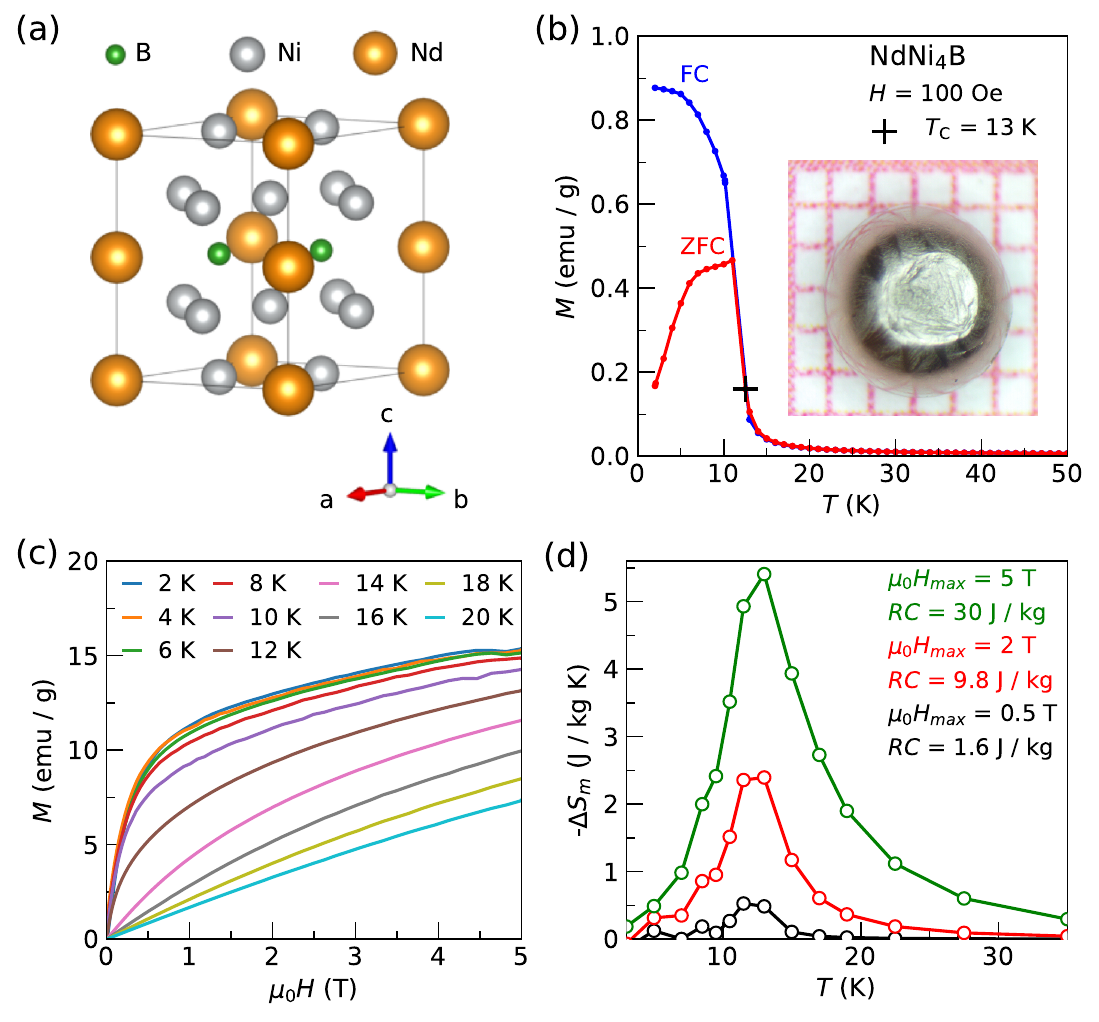}
  \caption{\label{fig:NICKEL}
    (a) Crystal structure of \ch{NdNi4B} illustrated by VESTA\cite{momma_vesta_2011, kuzma_x-ray_1983}. 
    (b) Magnetization of \ch{NdNi4B} showing the FM transition at $T_\text{C}=13$~K. Inset: a button of \ch{NdNi4B} produced by arc-melting. 
    (c) Several magnetization curves obtained near the transition temperature. 
    (d) Change in the magnetic entropy ($-\Delta S_m$) and refrigerant capacity ($RC$) of \ch{NdNi4B} near $T_\text{C}$.
  }
\end{figure}

The \ch{Nd$T$4B} family of materials crystallizes in the space group $P6$/$mmm$ comprising a layered structure shown in Fig.~\ref{fig:NICKEL}a for \ch{NdNi4B}~\cite{zlotea_determination_2002, salamakha_ndni4b_2003}. 
The bottom layer consists of a honeycomb $T$ lattice and a hexagonal Nd lattice. 
The second layer is then a kagome $T$ layer, followed by a third layer of a honeycomb B lattice interlacing a hexagonal Nd lattice (see also Fig.~S1). 
The fourth layer is then identical to the second and the fifth and final layer is identical to the first.
Both the kagome and honeycomb $T$ site are mixing sites wherein the full or partial substitution of one of Fe, Co, Ni for another leads to the high tunability of the material.
The magnetic moments order in an in-plane ferromagnetic orientation~\cite{burzo_rare_2023}. 

In this work, we synthesize the \ch{Nd$T$4B} compounds via a standard arc-melting procedure, which is ideal for manufacturability.
It produces polycrystalline samples, which allows us to ignore potential decreases in MCE size due to crystal orientation. 
We then investigate the effect of chemical substitution, investigating the change in peak temperature and size of the MCE. 
We show that, due to the highly tunable $T_C$, large MCE, and, in some select compositions, the appearance of a double transition due to multiple magnetic transitions, the \ch{Nd$T$4B} family is an ideal materials platform for the study of magnetocaloric refrigeration.

\section{\label{sec:methods}Experimental Methods}

\subsubsection{Material Synthesis}

We synthesized polycrystalline buttons of the \ch{Nd$T$4B}  materials via arc-melting~\cite{chen_ternary_1999}. 
Using an MRF SA-200 arc furnace, we melted stoichiometric quantities of Nd ingots (99.1\%) and B pieces (99.4\%) together with the desired combination of transition metals (Fe pieces (99.99\%), Co pieces (99.9\%), Ni shot (99.95\%)). 
We flipped and remelted the samples twice to promote homogeneity.
An example of one of the produced polycrystalline buttons is shown in the inset of Fig.~\ref{fig:NICKEL}b.

\subsubsection{Structural and Chemical Characterizations}

We confirmed the crystallographic structure using powder X-ray diffraction (PXRD) obtained by a Bruker D8 ECO instrument in the Bragg-Brentano geometry, using a copper source (Cu-K$_\alpha$) and a LYNXEYE XE 1D energy dispersive detector (Fig.~S2).
The crystal structure was refined using the FullProf suite~\cite{rodriguez-carvajal_recent_1993} and visualized using the VESTA program~\cite{momma_vesta_2011}.
We evaluated the chemical composition using energy dispersive X-ray spectroscopy (EDX) in an FEI Scios DualBeam electron microscope equipped with an Oxford detector (Table~S1, Fig.~S3, Fig.~S4).

\subsubsection{Magnetic Characterization}

We characterized the magnetic properties of our \ch{Nd$T$4B} materials as functions of both temperature and magnetic field using a Quantum Design 7 T MPMS 3 with vibrating sample magnetometry (VSM) and oven options. 
The magnitude of the MCE was characterized by determining the change in magnetic entropy ($\Delta S_m$) using the Maxwell equation~\cite{fruhling_characterization_2024,abramchuk_tuning_2019,kinami_magnetocaloric_2018}, 

\begin{equation}
\label{eq:maxwell}
\Delta S_m (T, \Delta H) = \int_0^{H_{\text{max}}} \left(\frac{\partial M}{\partial T} \right)_H dH
\end{equation}

Where $M$ is the magnetization, $T$ is the temperature and $H$ is the magnetic field. 
This equation was rewritten in the following form for the numerical evaluation of $\Delta S_m$ from discrete isotherms.

\begin{equation}
\label{eq:numerical}
    \begin{split}
        &\Delta S_m \left(T=\frac{T_1+T_2}{2}\right) = \\\left(\frac{1}{T_2-T_1}\right)&\times\left[ \int_0^{H_{\text{max}}} M(T_2,H)\, dH - \int_0^{H_{\text{max}}} M(T_1,H)\, dH \right]
    \end{split}
\end{equation}

where $T_1$ and $T_2$ are the temperatures of two neighboring isotherms, while the integration extends up to $H_\text{max}=0.5, 2,$ and $5$ T.

\section{\label{sec:results}Results and Discussion}

\subsubsection{MCE Metrics}

To quantify the MCE in the \ch{Nd$T$4B} system as necessary for the engineering of new refrigeration paradigms, we evaluated the MCE in terms of three standard metrics. 
The first one is the maximum change in magnetic entropy, $-\Delta S_{\text{MAX}}$.
This peak occurs near the onset of magnetic ordering and measures the maximum cooling effect of the material. 
Therefore, it is useful to first determine the Curie temperature $T_\text{C}$ as a function of temperature, as shown in Fig.~\ref{fig:NICKEL}b for \ch{NdNi4B}.
$T_\text{C}$ is determined from the minimum of the derivative of magnetization (Fig.~S5a,b).
From here, one needs isothermal magnetization curves, $M(H)$, at several temperatures around $T_\text{C}$ (Fig.~\ref{fig:NICKEL}c), which can be used to evaluate $-\Delta S_m$ in the vicinity of $T_\text{C}$ using Eq.~\ref{eq:numerical} (Fig.~\ref{fig:NICKEL}d). 
For \ch{NdNi4B} this maximum value is $-\Delta S_{\text {MAX}}=5.4$ \JKK~for a maximum field of 5 T. 

The second metric is the full width at half maximum (FWHM) of the entropy curve, $-\Delta S_m(T)$. 
This is a measure of the region over which the material has a significant cooling effect. 
In \ch{NdNi4B} (Fig.~\ref{fig:NICKEL}d), FWHM = 7 K at a maximum field of 5 T.

\begin{table*}
 \begin{tabular}{l|l|l|l|l}
 \hline
 \hline
 Composition & $T_\text{Peak}$ (K) & $-\Delta S_{\text {MAX}}$ (\JKK) & FWHM (K) & $RC$ (\JK)                  \\
 &  & 0.5 / 2 / 5 T & 0.5 / 2 / 5 T & 0.5 / 2 / 5 T\\
 \hline
 \ch{NdNi4B} & 13 & 0.53 / 2.4 / 5.4 & 3.7 / 5.0 / 7.3 & 1.6 / 9.8 / 30 \\
 \ch{NdCo4B} & 453 & 0.35 / 1.1 / 2.2 & 21 / 40 / 71 & 5.9 / 34 / 120  \\
 \ch{NdCo2Ni2B} & 130 & 0.30 / 0.78 / 1.5 & 15 / 36 / 150 & 3.4 / 21 / 170  \\
 \ch{NdCo3NiB} & 223 & 0.47 / 1.0 / 1.6 & 21 / 40 / 66 & 7.7 / 32 / 79  \\ 
 \ch{NdFeNi3B} & 111 & 0.10 / 0.43 / 1.1 & 260 / 290 / 290 & 18 / 93 / 250  \\
 \ch{NdFeCo3B} & 673 & 0.26 / 0.85 / 1.8 & 51 / 92 / 130 & 9.8 / 57 / 180  \\
 \ch{NdFe_{1.65}Co_{1.40}Ni_{0.95}B} & 519 & 0.15 / 0.59 / 1.4 & 78 / 130 / 160 & 8.5 / 54 / 160 \\
 \ch{NdFe_{1.15}Co_{0.46}Ni_{2.39}B} & 385 & 0.067 / 0.27 / 0.68 & 210 / 430 / 460 & 11 / 87 / 250 \\
 \hline
 \hline
 \end{tabular}
 \caption{\label{tab:METRICS}Comparing the MCE metrics in \ch{Nd$T$4B} materials, investigated here, for a maximum field of 0.5, 2, and 5 T.}
 \end{table*}

\begin{figure*}
\centering
  \includegraphics[width=\textwidth]{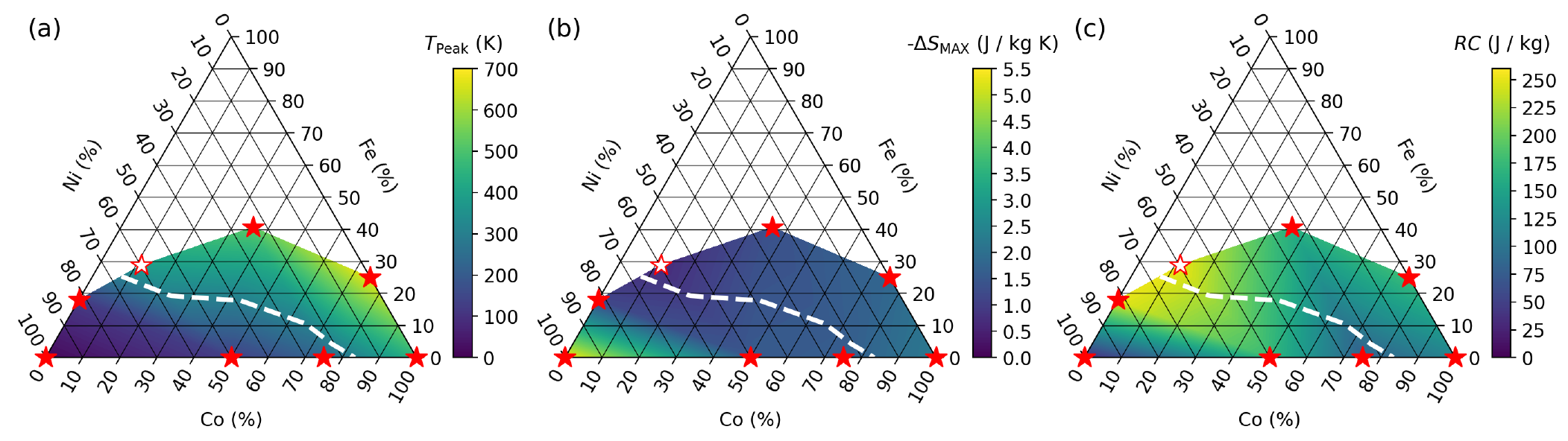}
  \caption{\label{fig:TERNARY}
    For various transition metal ratios, the color plots show the change in (a) temperature of the peak in $-\Delta S_m$, (b) the maximum change in the magnetic entropy at 5~T, and (c) the refrigerant capacity at 5~T. On all plots, the dashed white line represents a peak in $-\Delta S_m$ at 300 K.
  }
\end{figure*}

Finally, the refrigerant capacity ($RC$) is a measure of the total heat extracted over one thermodynamic cooling cycle. It is calculated by taking the integral of the entropy curve over the FWHM,
\begin{equation}
\label{eq:RC}
RC = \int_\text{FWHM}-\Delta S_m(T)dT
\end{equation}
This value is 30 \JK\ at 5~T  in \ch{NdNi4B} (Fig.~\ref{fig:NICKEL}d).

\subsubsection{MCE in \ch{NdT4B} Family}

We evaluated the above metrics for several \ch{Nd$T$4B} compositions (Fig.~S5, Fig.~S6) and summarized them in Table~\ref{tab:METRICS}. 
These values were used to construct the ternary diagrams in Fig.~\ref{fig:TERNARY} using a linear interpolation between the measured values.
The insolubility (white) regions in the ternary phase diagrams reflect the formation of impurity phases when the Fe content is $\gtrsim40\%$~\cite{gros_mossbauer_1988}.

The filled red stars in Fig.~\ref{fig:TERNARY} represent the initial 7 compositions of \ch{Nd$T$4B} measured in this work.
Using the color plots produced from these 7 compositions (Fig.~S7), we targeted a specific composition to engineer the maximum MCE, while ensuring that $-\Delta S_m(T)$ peaked near room temperature. 
This composition is marked by a white star in Fig.~\ref{fig:TERNARY}, corresponding to \ch{NdFe_{1.15}Co_{0.46}Ni_{2.39}B}.

To determine the ideal composition, it is instructive to examine the trends in our derived ternary phase diagrams. 
First, $T_\text{Peak}$ shown in Fig.~\ref{fig:TERNARY}a is highly sensitive to the choice of transition metal.
Increasing the Ni content quickly decreases both $T_\text{C}$ and $T_\text{Peak}$, while Co brings them closer to room temperature, and Fe brings them above room temperature.
We draw a dashed white line on the phase diagram, where $T_\text{Peak}$ is closer to 300~K for an ideal room temperature refrigerant material.
This line is overlaid on each panel of Fig.~\ref{fig:TERNARY}. 

Second, the maximum change in magnetic entropy, $-\Delta S_\text{MAX}$, in Fig.~\ref{fig:TERNARY}b is at its largest value in undoped \ch{NdNi4B}. 
$-\Delta S_\text{MAX}$ takes its second largest value in undoped \ch{NdCo4B}.
We then see that any mixing of the transition metals reduces $-\Delta S_\text{MAX}$ from the value in single transition metal materials.  

The third trend is that while $-\Delta S_\text{MAX}$ shows mild changes across the various mixed transition metal \ch{Nd$T$4B}, the refrigerant capacity ($RC$) shows significant changes. 
The reason for the variance of $RC$ is a considerable change in the width of the magnetic transition when all three transition metals are mixed.
As a result, the FWHM of the $-\Delta S_m(T)$ curve increases, which enhances the $RC$ despite a reduced $-\Delta S_\text{MAX}$.
Because achieving a $T_\text{Peak}$ near 300 K requires a significant mixing of different transition metals, we choose to focus on maximizing $RC$. 
This means maximizing the temperature range over which a material is consistently capable of cooling while ignoring its maximum cooling ability. 
Choosing the maximum predicted $RC$ along the 300 K line led to predicting the composition \ch{NdFe_{1.20}Co_{0.48}Ni_{2.32}B}.

\subsubsection{Engineered Material}

\begin{figure*}
\centering
  \includegraphics[width=\textwidth]{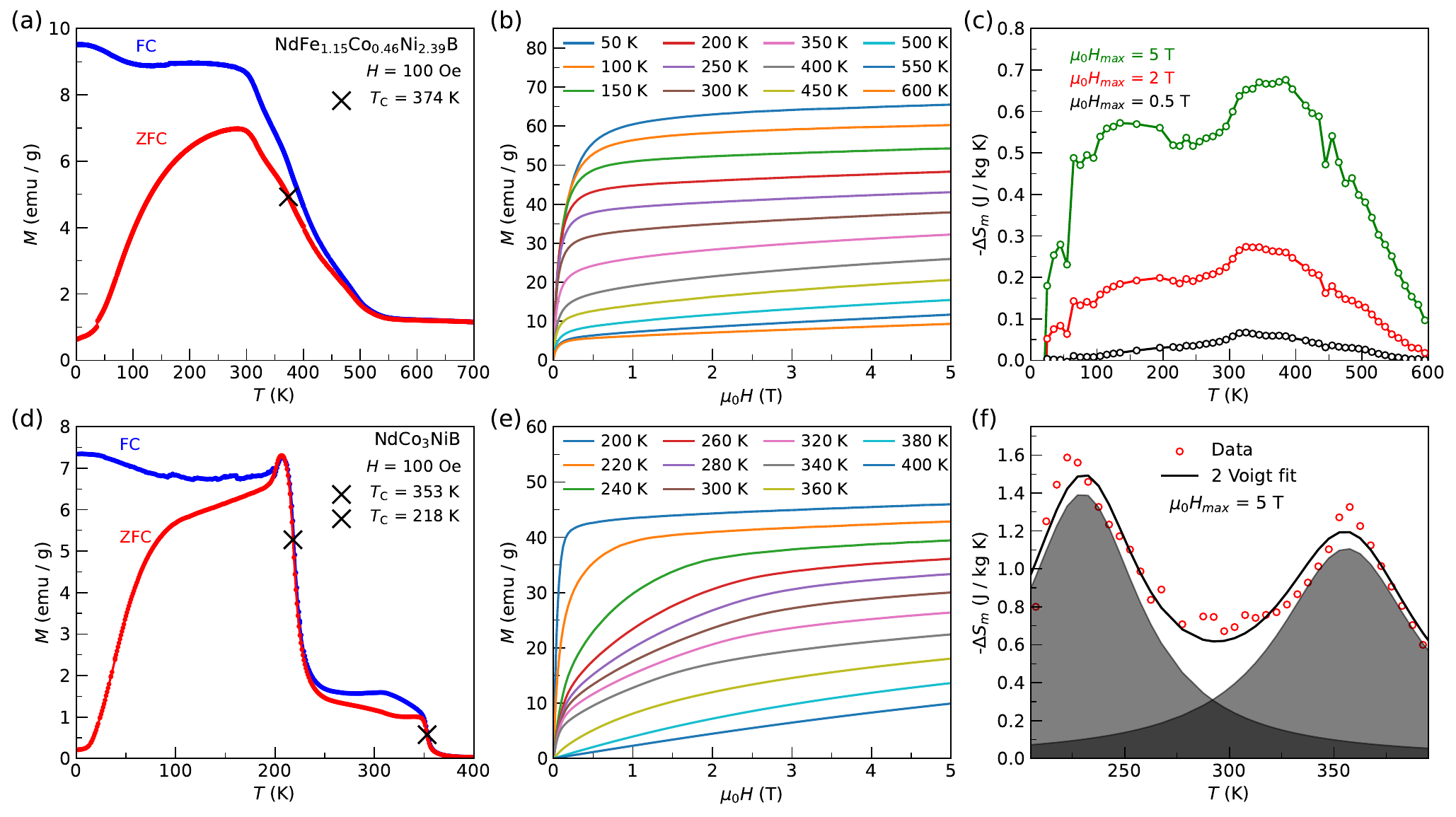}
  \caption{\label{fig:EXAMPLES}
    Top: characterizing the MCE in \ch{NdFe_{1.15}Co_{0.46}Ni_{2.39}B}. (a) Magnetization as a function of temperature showing the FM transition. (b) Magnetization isotherms over a wide temperature range. (c) $-\Delta S_m$ over the wide FM transition. 
    Bottom: characterizing the MCE in \ch{NdCo3NiB}. (d) Two FM transitions in \ch{NdCo3NiB}. (e) Magnetic isotherms over a wide temperature range. Two regions of increased separation between neighboring isotherms can be seen around 220 K and 360 K. (f) A two-peak behavior in $-\Delta S_m$ fitted to the sum of two Voigt functions centered around 230~K and 356~K.
  }
\end{figure*}

\begin{table*}
 \begin{tabular}{l|l|l|l|l}
 \hline
 \hline
 Material & $T_\text{Peak}$ (K) & $-\Delta S_{\text {MAX}}$ (\JKK) & FWHM (K) & $RC$ (\JK) \\
 \hline
 \ch{NdCo3NiB} & 230 / 356 & 1.4 / 1.1 & 66 / 80 & 73 / 80 \\
 \ch{NdCo2Ni2B} & 60 / 137 & 0.77 / 1.3 & 51 / 120 & 32 / 120 \\
 \ch{NdFeNi3B} & 94 / 238 & 0.89 / 0.70 & 130 / 150 & 91 / 140 \\
 \ch{NdFe_{1.15}Co_{0.46}Ni_{2.39}B} & 122 / 359 & 0.47 / 0.67 & 140 / 310 & 52 / 170 \\
 \hline
 \hline
 \end{tabular}
 \caption{\label{tab:MULTISTAGE}Comparing the MCE metrics in \ch{Nd$T$4B} materials with two-peak behavior for a maximum magnetic field of 5~T. 
 The listed pairs of values correspond to the low/high temperature peaks.
 The values reported in this Table come from the Voigt fits (Eq.~\ref{eq:VOIGT_FIT}), whereas the values in Table~\ref{tab:METRICS} are from the unfitted data.}
 \end{table*}

The experimental composition of the engineered system, determined by EDX, deviated slightly from the nominal composition, yielding a value of \ch{NdFe_{1.15}Co_{0.46}Ni_{2.39}B}. 
The magnetization as a function of temperature for \ch{NdFe_{1.15}Co_{0.46}Ni_{2.39}B} is shown in Fig.~\ref{fig:EXAMPLES}a. 
The wide magnetic transition is consistent with high levels of disorder due to mixing three transition metals (Fe, Co, and Ni).
This leads to a relatively consistent MCE over a wide temperature range. 
The magnetization curves as a function of field for multiple temperatures are shown in Fig.~\ref{fig:EXAMPLES}b. 
The separation between neighboring magnetization isotherms remains relatively constant, indicating a consistent MCE over a large temperature range. 
Also, the saturation moment continues to increase over a range of hundreds of Kelvin below $T_C$. 
These two factors lead to the broad $-\Delta S_m(T)$ curve observed in Fig.~\ref{fig:EXAMPLES}c.
As expected, $-\Delta S_\text{MAX}$ is small; in fact, it is the lowest value in the \ch{Nd$T$4B} family at 0.68 \JKK for a maximum field of 5~T. 

The peak in the MCE of \ch{NdFe_{1.15}Co_{0.46}Ni_{2.39}B} (Fig.~\ref{fig:EXAMPLES}c) occurs at a slightly higher temperature than hoped for (385 K instead of 300 K).
In this composition, $-\Delta S_m$ remains close to its maximum value throughout a wide temperature range, leading to a wide FWHM of 457~K, the widest in the series. 
The remarkably large FWHM, despite a small $\Delta S_\text{MAX}$, yields a sizable $RC$ of 250 \JK. 
As such, \ch{NdFe_{1.15}Co_{0.46}Ni_{2.39}B}, while not a particularly strong candidate for magnetic refrigeration at any given temperature, is a consistent magnetocaloric material over a wide temperature range. 
This is a desirable quality to engineer and can be the focus of future works as it enables consistent refrigeration with a single material over a wide temperature range. 
Other materials such as \ch{Eu4PdMg} and \ch{Gd3Ni6AlGa} also show a broad MCE~\cite{li_reversible_2014, aseguinolaza_broad_2025}.
They have a smaller FWHM between 100 and 200~K, but a larger $-\Delta S_\text{MAX}$, resulting in larger $RC$ of 832 \JK for \ch{Eu4PdMg} and 560 \JK for \ch{Gd3Ni6AlGa}, albeit near 100~K instead of room temperature. 
These materials are therefore better examples of overall cooling power, while our material provides an example of a wide cooling range for room temperature applications.

\subsubsection{Two Stage Cooling Behavior}

An interesting feature in a few mixed transition metal \ch{Nd$T$4B} materials is the presence of two peaks in $-\Delta S_m(T)$.
For example, the wide $-\Delta S_m(T)$ curve in the targeted material \ch{NdFe_{1.15}Co_{0.46}Ni_{2.39}B} (Fig.~\ref{fig:EXAMPLES}c) is made of two peaks, which are best resolved at 5~T (green curve).
A more obvious example is shown in Figs.~\ref{fig:EXAMPLES}d-f for \ch{NdCo3NiB}, where two FM transitions appear in the magnetization data (Fig.~\ref{fig:EXAMPLES}d) at 218~K and 353~K. 
This leads to two temperature regimes with increased separation between neighboring isotherms in Fig.~\ref{fig:EXAMPLES}e, corresponding to the two peaks at $T_\text{Peak}=$ 223~K and 358~K in $-\Delta S_m(T)$ evaluated for $H_\text{max}=5$~T in Fig.~\ref{fig:EXAMPLES}f.

The two-peak MCE has been reported previously in a few rare-earth-rich materials such as \ch{HoIn2}, \ch{Dy2Cu2Cd}, \ch{Ho2Cu2Cd}, and \ch{Gd$_{1-x}$Ho$_x$Ni} but at lower temperatures ($T<100$~K) and in narrower temperature ranges~\cite{zhang_magnetocaloric_2009,zhang_review_2019, yi_large_2017, jiang_double_2018, zhang_excellent_2016}. 
In these materials, the peaks correspond to the initial onset of magnetic ordering, followed by a spin reorientation that results in another rapid change in magnetic entropy.
Our results in Fig.~\ref{fig:EXAMPLES}d-f reveal such a behavior near room temperature and in a wider temperature range.
Materials with a two-peak MCE can be leveraged for multi-stage cooling, circumventing the need for multiple magnetocaloric materials. 

For a quantitative analysis of the two-peak behavior in Fig.~\ref{fig:EXAMPLES}f, we fit each magnetocaloric peak to a Voigt profile, which is a convolution of a Lorentzian and a Gaussian,

\begin{equation}
    \label{eq:VOIGT}
    V(x;\sigma,\gamma) = \int^{\infty}_{-\infty}G(x';\sigma)L(x-x';\gamma)dx'
\end{equation}

where $G$ is a Gaussian profile of standard deviation $\sigma$ and $L$ is a Lorentzian profile of scaling factor $\gamma$. 
The integration parameter $x$ is the temperature.
The entropy data are then fit to the sum of two Voigt profiles,

\begin{equation}
    \label{eq:VOIGT_FIT}
    -\Delta S_M = A_1V(x-\mu_1;\sigma_1,\gamma_1) + A_2V(x-\mu_2;\sigma_2,\gamma_2)
\end{equation}

where $A_1$/$A_2$ is a multiplicative factor to match the peak amplitudes and $\mu_1$/$\mu_2$ corresponds to $T_\text{Peak}$ for each curve. 
For \ch{NdCo3NiB}, this results in low temperature values; $T_{\text{Peak}} = 230$ K, $-\Delta S_{\text{MAX}} = 1.4$ \JKK and $RC = 73$ \JK, and high temperature values; $T_{\text{Peak}} = 356$ K, $-\Delta S_{\text{MAX}} = 1.1$ \JKK and $RC = 80$ \JK. 
The comparable values of $RC$ signify comparable refrigerant capacity at both stages of cooling -- an ideal property for a two-stage magnetocaloric material.

We performed this analysis on other \ch{Nd$T$4B} materials with a two-peak MCE (Fig.~S8). 
The samples were also characterized using PXRD and EDX to confirm that the two-stage cooling did not arise from phase separation within the sample (Fig.~S4). 
The results are summarized in Table~\ref{tab:MULTISTAGE}. 
In \ch{NdCo3NiB} and \ch{NdCo2Ni2B} specifically, we see the ability to engineer each curve separately. 
Increasing the amount of Co increases the $T_\text{Peak}$ values for each curve; however, the values of $-\Delta S_\text{MAX}$ and $RC$ change unequally. 
Future efforts could focus on engineering two-stage cooling in other materials to maximize the size of the MCE at various temperatures, tailored to specific technological needs.

\section{\label{sec:conclusions}Conclusions}

\ch{Nd$T$4B} is a highly tunable magnetic material family. This high degree of tunability enables engineering of the magnetocaloric effect through specific transition-metal mixing on the $T$ site.
This is reflected in the ternary phase diagrams in Fig.~\ref{fig:TERNARY}.
Using a linear interpolation between seven representative samples, we were able to target a composition for maximum $RC$ near room temperature. 
The engineered composition, \ch{NdFe_{1.15}Co_{0.46}Ni_{2.39}B}, has a peak MCE at 385~K, and shows the largest $RC$ in the \ch{Nd$T$4B} system, significantly larger than \ch{NdCo3NiB}, the only other material with a $T_\text{Peak}$ close to room temperature. 

This approach to engineering the ideal magnetocaloric composition within a material family, by constructing a phase diagram from a few representative samples and predicting the ideal one through a linear interpolation, has proven effective in the case of \ch{Nd$T$4B}, which contains only three transition-metal elements (Fe, Co, and Ni).
We note that higher-dimensional phase diagrams could be applied to more well-studied materials such as the high-entropy transition metal alloys; they could also be used to guide the machine learning approaches already being applied to search for promising magnetocaloric materials~\cite{zhang_tunable_2014, guo_large_2022, qiao_giant_2025}.

We also found that in materials with two peaks in the magnetic entropy curve, the behavior of each curve changes somewhat independently and are thus, potentially, independently tunable. 
This effect can be a subject of future investigations for engineering magnetocaloric materials for multi-stage cooling.

\section*{ACKNOWLEDGMENTS}
The work at Boston College (magnetization, EDX, and X-ray characterizations) was funded by the U.S. Department of Energy, Office of Basic Energy Sciences, Division of Physical Behavior of Materials under award number DE-SC0023124.
This material is based upon work supported by the Air Force Office of Scientific Research (AFOSR) under award number FA9550-23-1-0124, AFOSR LRIR 23RXCOR003, AFOSR LRIR 26RXCOR010, AFOSR LRIR 24RQCOR004, internal funding from the Air Force Research Laboratory's Aerospace Systems Directorate, and support from the joint NSF-AFOSR INTERN Program.




\bibliography{Bibliography}

\end{document}



\title{Supplemental Material:\\Engineering the Magnetocaloric Effect in Nd$T_4$B}

\author{Kyle~W.~Fruhling}
\email{fruhling@bc.edu}
\affiliation{Department of Physics, Boston College, Chestnut Hill, MA 02467, USA}

\author{Enrique~O.~González~Delgado}
\affiliation{Department of Physics, Boston College, Chestnut Hill, MA 02467, USA}

\author{Siddharth~Nandanwar}
\affiliation{Department of Physics, Boston College, Chestnut Hill, MA 02467, USA}

\author{Xiaohan~Yao}
\affiliation{Department of Physics, Boston College, Chestnut Hill, MA 02467, USA}

\author{Zafer~Turgut}
\affiliation{Aerospace Systems Directorate, Air Force Research Laboratory, Wright Patterson Air Force Base, OH 45433, USA}

\author{Michael~A.~Susner}
\affiliation{Materials and Manufacturing Directorate, Air Force Research Laboratory, Wright Patterson Air Force Base, OH 45433, USA}

\author{Fazel~Tafti}
\email{fazel.tafti@bc.edu}
\affiliation{Department of Physics, Boston College, Chestnut Hill, MA 02467, USA}

\maketitle


\section{Crystal Structure}

\begin{figure*}
\centering
  \includegraphics[width=\textwidth]{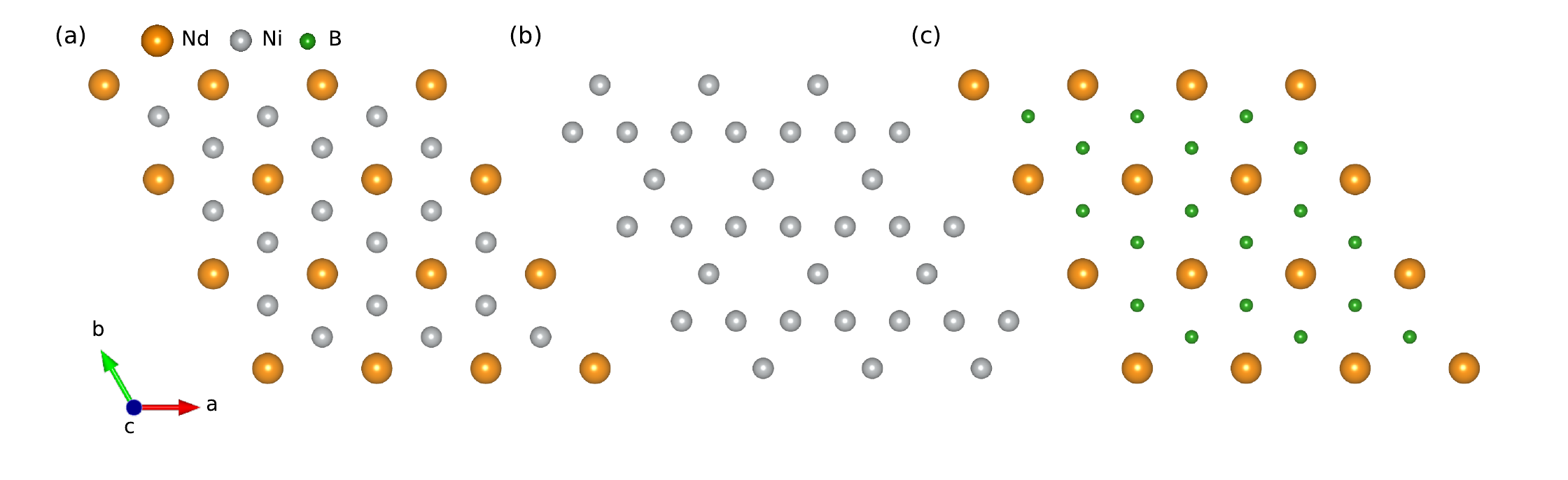}
  \caption{\label{fig:LAYERS}
    Layers of \ch{NdNi4B} (a) Nd/Ni hexagonal/honeycomb layer, (b) Ni Kagome layer, and (c) Nd/B hexagonal/honeycomb layer.
  }
\end{figure*}

As described in the main text, \ch{Nd$T$4B} crystallizes in the space group $P6/mmm$ and consists of a few layers. 
The bottom layer is a Nd-$T$ layer with a hexagonal Nd layer interlaced with a honeycomb $T$ layer (Fig.~\ref{fig:LAYERS}a). 
The second layer is a pure $T$ layer with a kagome structure (Fig.~\ref{fig:LAYERS}b). 
The third layer is a Nd-B layer with a triangular Nd lattice and a boron honeycomb lattice (Fig.~\ref{fig:LAYERS}c). 
The fourth layer is then identical to the second layer and the fifth layer is identical to the first.

The structure of \ch{NdNi4B} has also been refined in the space group $Imma$ with $a'=a$, $b'=c$, and $c'=3a\sqrt{3}$ where $a$, $b$, and $c$ are the lattice parameters in the $P6/mmm$ space group~[26]. 
This structure has not been reported for either \ch{NdCo4B} or any of the previously reported mixed transition metal compounds. 
Since we can dope Ni with Co and Fe, we believe the correct structure must be $P6/mmm$ because it would not be feasible to dope Ni with Co and Fe while the space group changes between the parent compounds. 

\section{PXRD Analysis}

\begin{figure*}
\centering
  \includegraphics[width=0.6\textwidth]{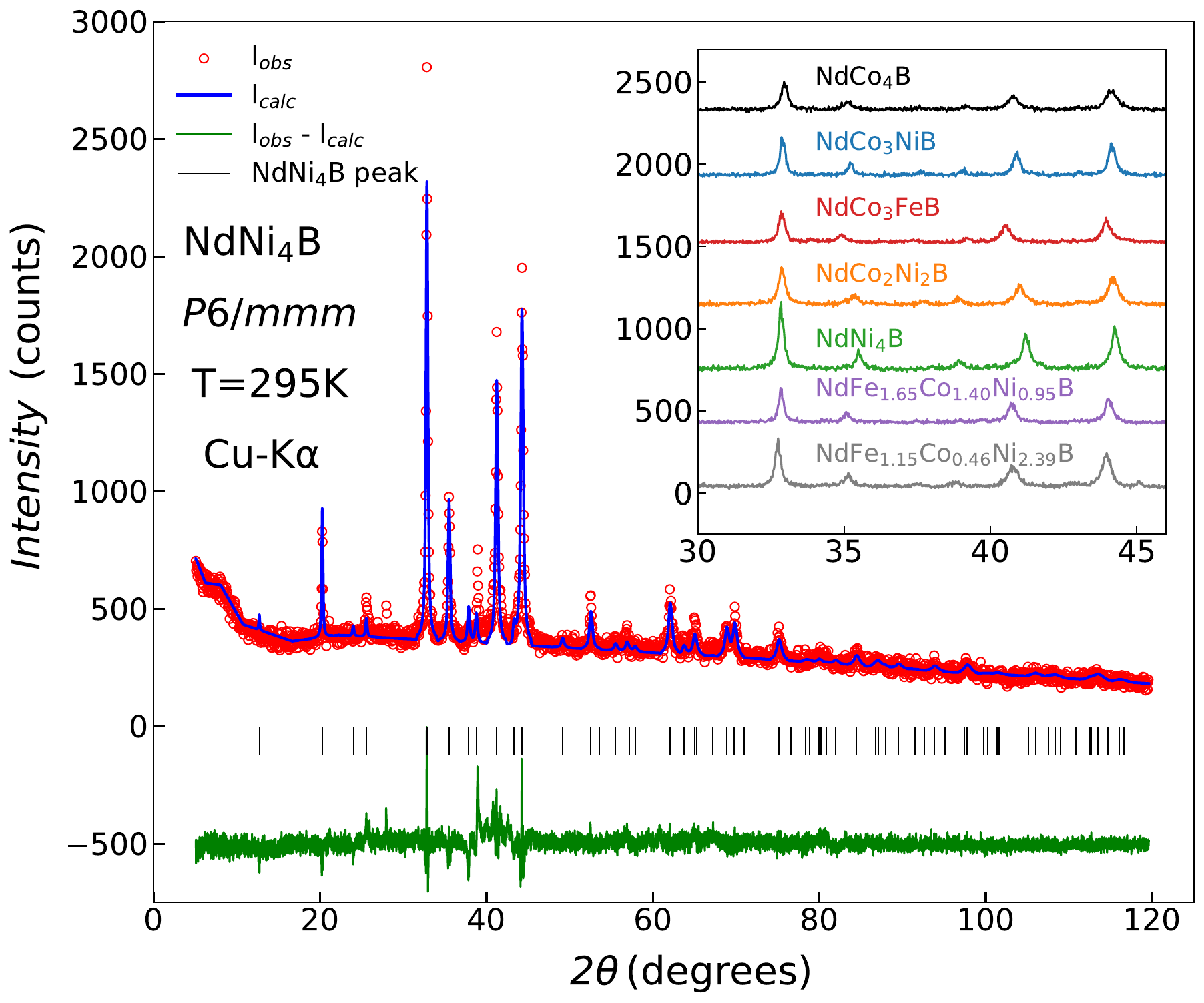}
  \caption{\label{fig:REFINEMENT}
    All \ch{Nd$T$4B} materials crystallize in the same structure as the parent compound \ch{NdNi4B}. The main panel shows a Rietveld refinement on the powder X-ray diffraction pattern of the parent compound \ch{NdNi4B}. The inset shows the systematic change of Bragg peaks in the diffraction patterns of other compositions.
  }
\end{figure*}

The crystal structure of \ch{Nd$T$4B} was confirmed, as well as the lack of any significant competing phases, using Powder X-Ray Diffraction (PXRD). 
Figure~\ref{fig:REFINEMENT} shows the Rietveld refinement of the X-ray pattern of \ch{NdNi4B} in space group $P6/mmm$.
The inset shows the systematic change of Bragg peaks for other compositions of \ch{Nd$T$4B}. 

\section{EDX Analysis}

The chemical compositions of \ch{Nd$T$4B} samples were confirmed by energy dispersive x-ray spectroscopy (EDX) using a scanning electron microscope (SEM).
The electron beam had a current of 0.4~nA and was accelerated under a voltage of 20~keV.
Different areas were scanned on each sample to calculate the average and standard deviation of the atomic percentage in each sample. 
Since boron has too small of a scattering cross section for EDX, its atomic percentage is reported as the nominal amount.
The final compositions with errors are reported in Tab.~\ref{tab:COMPOSITIONS}.

\begin{figure*}
\centering
  \includegraphics[width=.6\textwidth]{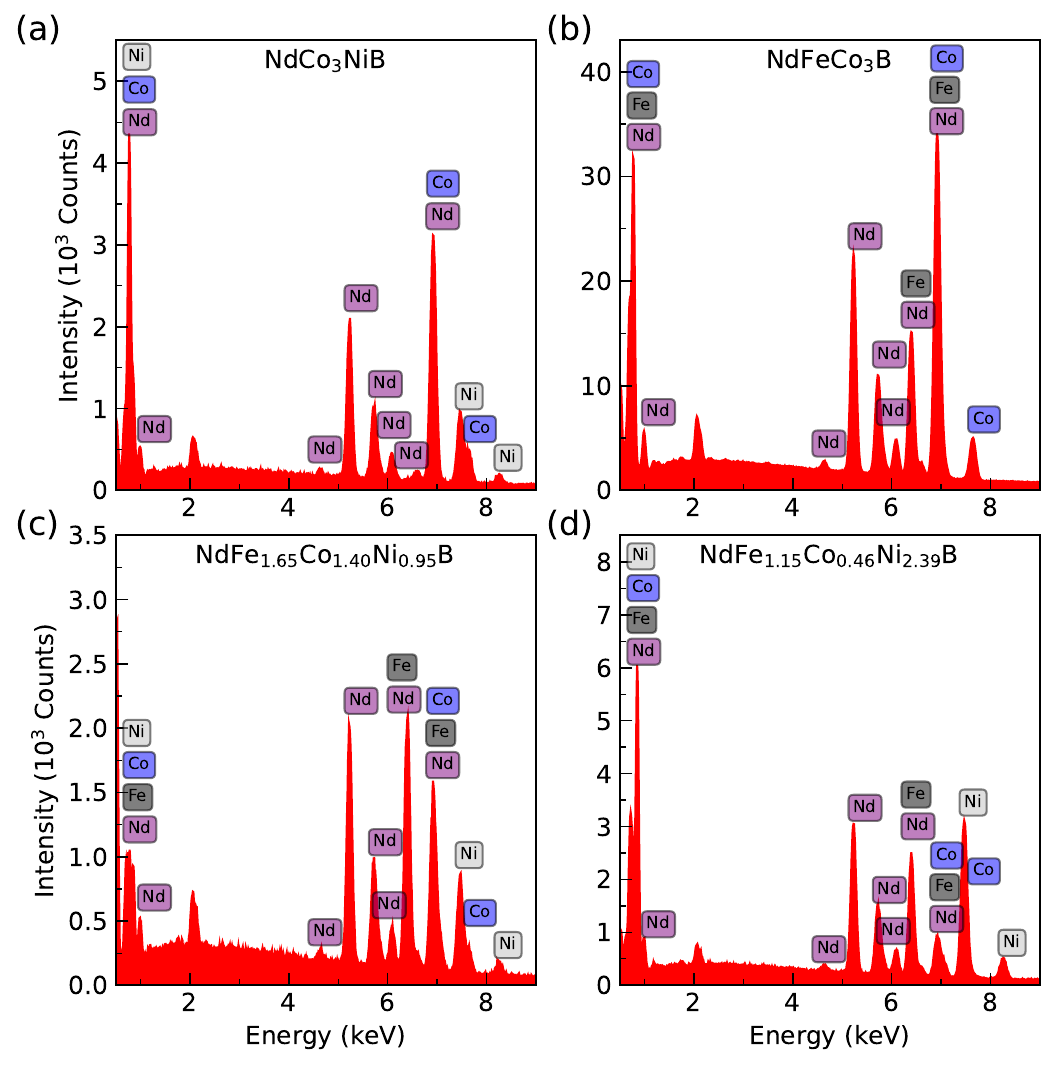}
  \caption{\label{fig:EDX}
    EDX spectra for (a) \ch{NdCo3NiB}, (b) \ch{NdFeCo3B}, (c) \ch{NdFe$_{1.65}$Co$_{1.40}$Ni$_{0.95}$B}, and (d) \ch{NdFe$_{1.15}$Co$_{0.46}$Ni$_{2.39}$B}. The persistent peak near 2 keV is from a 5~nm layer of Pt coating on the samples necessary for the EDX measurements.
  }
\end{figure*}

\begin{figure*}
\centering
  \includegraphics[width=1\textwidth]{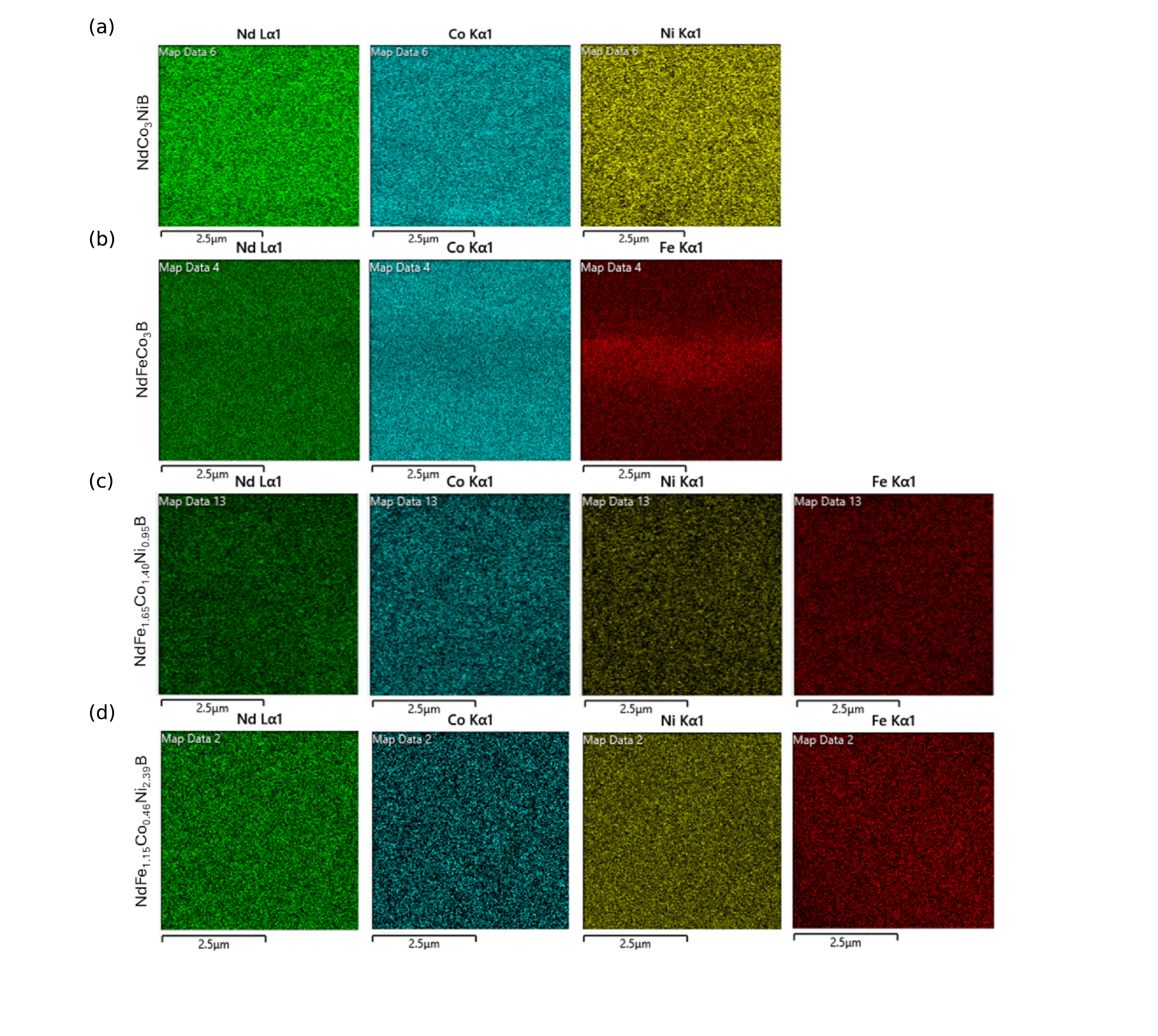}
  \caption{\label{fig:COLORMAPS}
    EDX colormaps for (a) \ch{NdCo3NiB}, (b) \ch{NdFeCo3B}, (c) \ch{NdFe$_{1.65}$Co$_{1.40}$Ni$_{0.95}$B}, and (d) \ch{NdFe$_{1.15}$Co$_{0.46}$Ni$_{2.39}$B}.
  }
\end{figure*}

The EDX spectra for four representative \ch{Nd$T$4B} samples are presented in Fig.~\ref{fig:EDX}.
Additionally, color maps of four compositions are shown in Fig.~\ref{fig:COLORMAPS} to examine phase homogeneity within each sample. 
We did not find microstructures in any of the samples, except in \ch{NdFeCo3B}, where some inhomogeneity in the distribution of iron was found. 
Despite this, \ch{NdFeCo3B} does not show two clear transitions or any two-peak behavior in the MCE, and other two-peak materials like \ch{NdCo3NiB} do not show any phase separation. This makes it clear that the two-peak behavior is not a result of phase separation.

\section{MCE Analysis}

\begin{figure*}
\centering
  \includegraphics[width=1\textwidth]{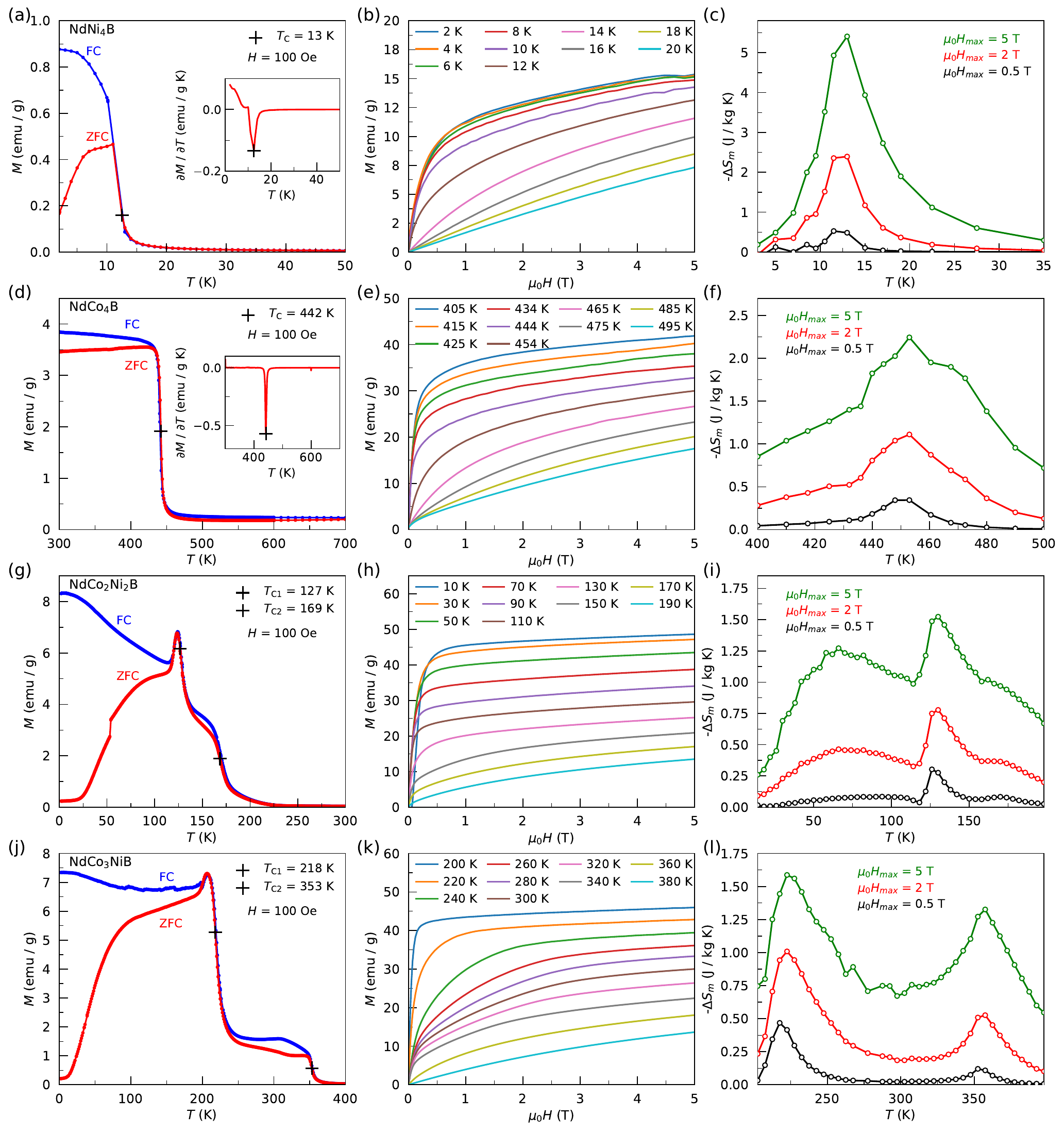}
  \caption{\label{fig:CHARACTERIZE_1}
    Magnetization vs. temperature, magnetization vs. field, and $-\Delta S_m$ as a function of temperature for (a-c) \ch{NdNi4B}, (d-f) \ch{NdCo4B}, (g-i) \ch{NdCo2Ni2B}, and (j-l) \ch{NdCo3NiB}.
  }
\end{figure*}

\begin{figure*}
\centering
  \includegraphics[width=1\textwidth]{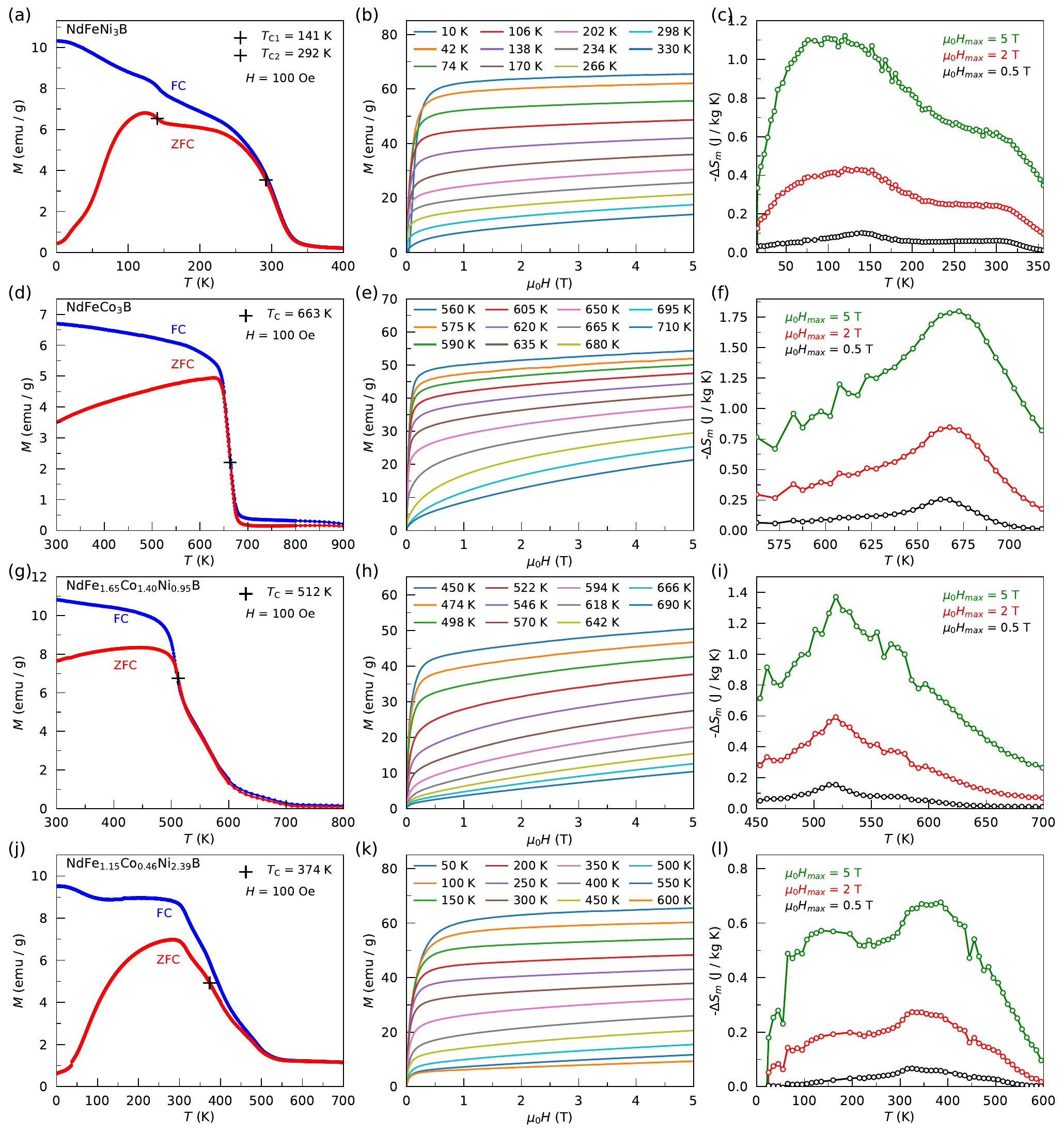}
  \caption{\label{fig:CHARACTERIZE_2}
    Magnetization vs. temperature, magnetization vs. field, and $-\Delta S_m$ as a function of temperature for (a-c) \ch{NdFeNi3B}, (d-f) \ch{NdFeCo3B}, (g-i) \ch{NdFe$_{1.65}$Co$_{1.40}$Ni$_{0.95}$B}, and (j-l) \ch{NdFe$_{1.15}$Co$_{0.46}$Ni$_{2.39}$B}.
  }
\end{figure*}

\begin{figure*}
\centering
  \includegraphics[width=1\textwidth]{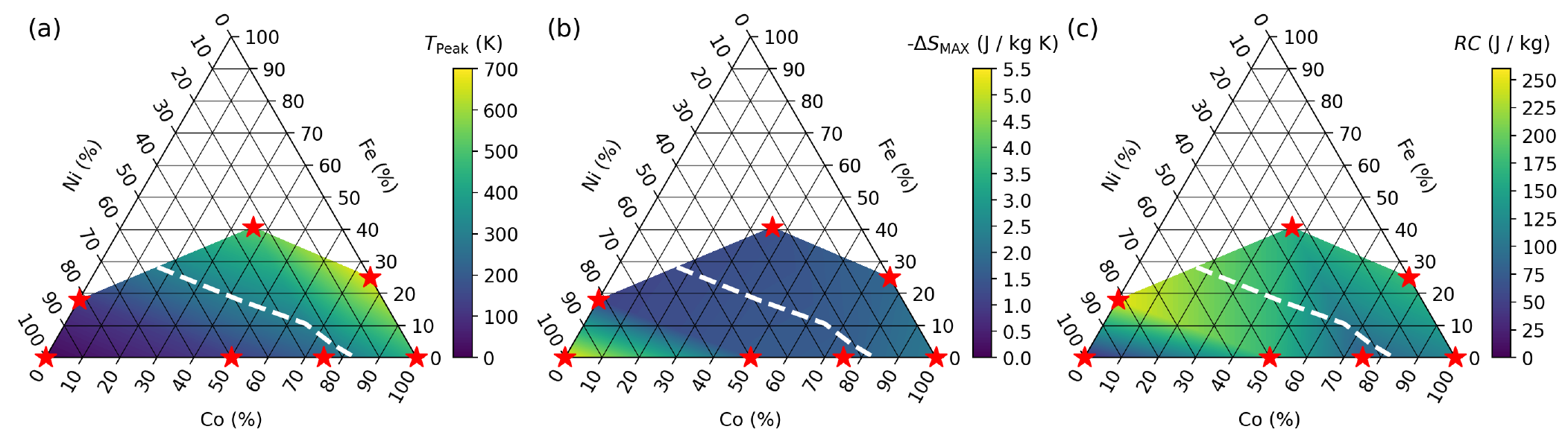}
  \caption{\label{fig:TERNARY_ORIGINALS}
    Ternary phase diagrams of the first seven \ch{Nd$T$4B} materials made, showing: (a) temperature of the peak in $-\Delta S_M$, (b) the maximum of the change in magnetic entropy at 5 T, and (c) the refrigerant capacity at 5 T. On all plots, the dashed white line represents a peak in $-\Delta S_M$ at 300 K.
  }
\end{figure*}

Figures~\ref{fig:CHARACTERIZE_1} and \ref{fig:CHARACTERIZE_2} show the magnetic and magnetocaloric characterizations of the studied materials. 
To determine the transition temperature around which to take isotherms, we first measured the moment as a function of temperature and determined the Curie temperature $T_C$ from the minimum of the $\partial M/\partial T$ curve, as shown in the insets of Fig.~\ref{fig:CHARACTERIZE_1}a and Fig.~\ref{fig:CHARACTERIZE_1}b.
Then, we measured several $M(H)$ isotherms, which were used to calculate the change in magnetic entropy with temperature. 
Using these entropy curves, we were able to extract our three MCE metrics: $T_{\text{Peak}}$, $\Delta S_{\text{MAX}}$, and $RC$. 
We then compiled the ternary plots shown in Fig.~\ref{fig:TERNARY_ORIGINALS}. 
Using these colormaps we decided to target the composition \ch{NdFe_{1.20}Co_{0.48}Ni_{2.32}B} in order to shift $T_{\text{Peak}}$ to near 300 K and maximize $RC$.

\subsubsection{Two-Peak Behavior}

\begin{figure*}
\centering
  \includegraphics[width=.6\textwidth]{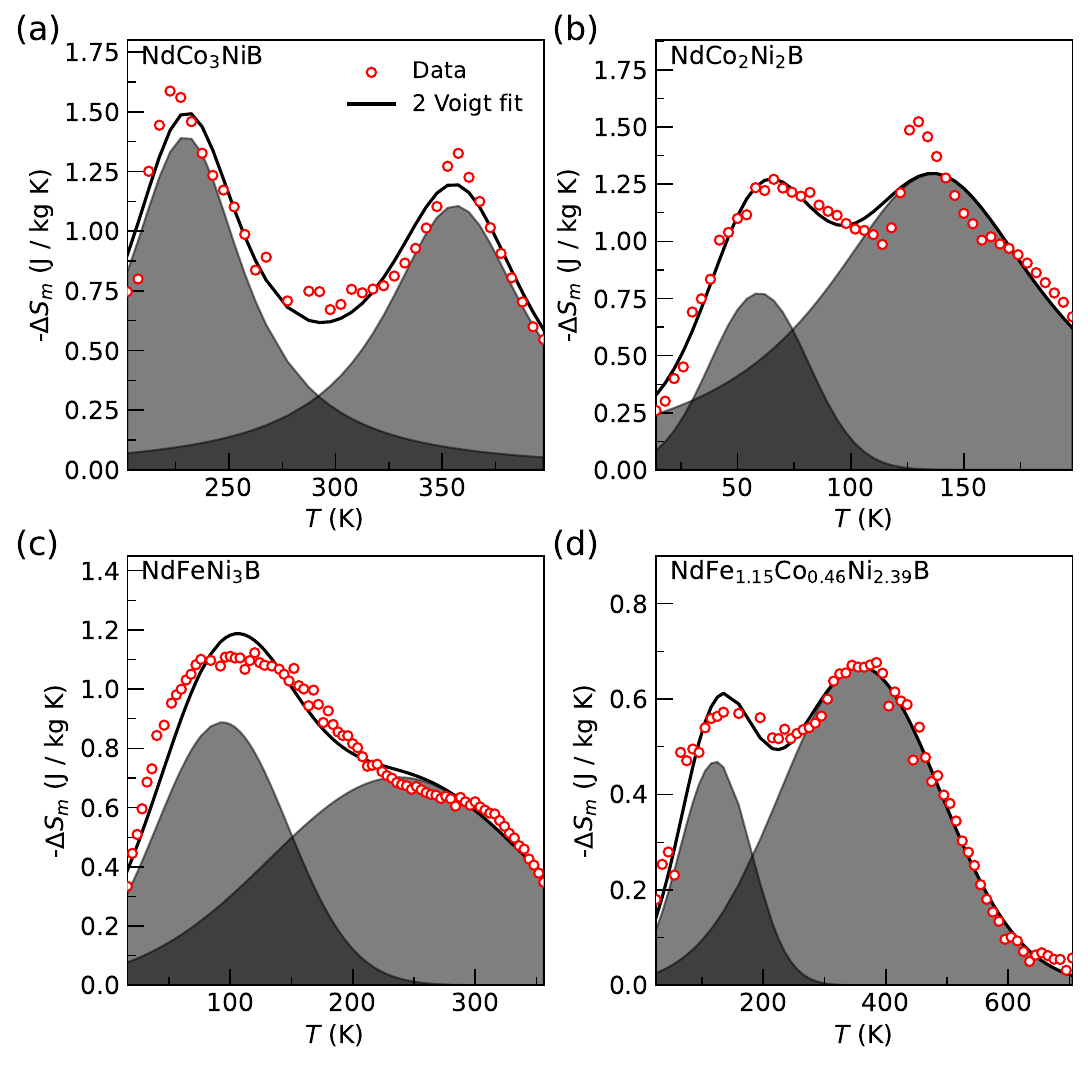}
  \caption{\label{fig:TWO_PEAKS}
    Magnetic entropy plots with Voigt fits for (a) \ch{NdCo3NiB}, (b) \ch{NdCo2Ni2B}, (c) \ch{NdFeNi3B}, (d) \ch{NdFe$_{1.15}$Co$_{0.46}$Ni$_{2.39}$B}. The red open dots represent the measured values, while the solid black line represents the two-Voigt fit discussed in the main text. The shaded curves are the individual Voigt profiles that make up the two-Voigt fit. For all these plots, the $-\Delta S_m$ values are calculated for a maximum field of 5 T.
  }
\end{figure*}

As mentioned in the main text, some \ch{Nd$T$4B} materials show two transitions or one very wide transition leading to $-\Delta S_m$ plots with two peaks. 
In Fig.~\ref{fig:TWO_PEAKS}, we present the magnetic entropy data for the materials with clear two-peak behavior. 
We also show the two-Voigt profile fit as described in Eq.~5 in the main text. 
Under this fit, we separate the two Voigt profiles into separate curves shown as the shaded regions. 
From these individual curves, we extract the MCE metrics reported in Table II in the main text.

\begin{table*}
 \begin{tabular}{l|l}
 \hline
 \hline
 Targeted Composition & Measured Composition \\
 \hline
 \ch{NdNi4B} & \ch{Nd_{1.0321(9)}Ni4B}\\
 \ch{NdCo4B} & \ch{Nd_{1.05(2)}Co4B}\\
 \ch{NdCo2Ni2B} & \ch{Nd_{1.029(2)}Co_{2.01(3)}Ni_{1.99(3)}B}\\
 \ch{NdCo3NiB} & \ch{Nd_{1.05(1)}Co_{2.99(2)}Ni_{1.01(2)}B}\\
 \ch{NdFeCo3B} & \ch{Nd_{1.010(3)}Fe_{1.04(3)}Co_{2.96(3)}B} \\
 \ch{NdFeNi3B} & \ch{Nd_{1.04(2)}Fe_{0.72(2)}Ni_{3.28(2)}B} \\
 \ch{NdFe_{1.33}Co_{1.33}Ni_{1.33}B} & \ch{Nd_{0.97(2)}Fe_{1.64(4)}Co_{1.40(3)}Ni_{0.95(2)}B}\\
 \ch{NdFe_{1.20}Co_{0.48}Ni_{2.32}B} & \ch{Nd_{1.01(2)}Fe_{1.15(5)}Co_{0.46(1)}Ni_{2.39(4)}B} \\
 \hline
 \hline
 \end{tabular}
 \caption{\label{tab:COMPOSITIONS}EDX results for compositions investigated in this study.}
 \end{table*}

